\documentclass[%
prl,%
reprint,%
twocolumn,%
superscriptaddress,%
footinbib,
aps,%
10pt]{revtex4-1}
\usepackage{amsmath}
\usepackage{hyperref}
\hypersetup{colorlinks=true,citecolor=blue,linkcolor=blue,filecolor=blue,urlcolor=blue,breaklinks=true}
\usepackage{graphicx}
\usepackage{amsfonts}
\usepackage{mathrsfs} 
\usepackage{amsthm}
\usepackage{amsmath}
\usepackage{amssymb}
\usepackage{bbm} 
\usepackage{mathtools} 
\usepackage[table]{xcolor} 
\usepackage{tikz}
\usepackage{pifont}     
\usepackage{enumitem}   
\newcommand{\ket}[1]{\vert #1 \rangle}%
\newcommand{\bra}[1]{\langle #1 \vert}%
\newcommand\proj[1]{\vert #1 \rangle \langle #1 \vert}%
\newcommand{\opn}[1]{\operatorname{#1}}%
\DeclareMathOperator{\tr}{Tr}
\newcommand{\1}{\mathbbm{1}}%

\newcommand*{\cH}{\mathcal{H}}
\newcommand*{\cT}{\mathcal{T}}
\newcommand*{\bE}{\mathbb{E}}

\newcommand{\sectionprl}[1]{{\par\it#1.---}}

\begin{document}

\title{Optimal Verification of Two-Qubit Pure States}%

\author{Kun Wang}%
\affiliation{Department of Computer Science and Technology, State Key Laboratory for Novel Software Technology, %
Nanjing University, Nanjing, 210093, China}%
\affiliation{Shenzhen Institute for Quantum Science and Engineering, %
Southern University of Science and Technology, Shenzhen, 518055, China}

\author{Masahito Hayashi}%
\email{masahito@math.nagoya-u.ac.jp}%
\affiliation{Graduate School of Mathematics, Nagoya University, Nagoya, 464-8602, Japan}%
\affiliation{Shenzhen Institute for Quantum Science and Engineering, %
Southern University of Science and Technology, Shenzhen, 518055, China}%
\affiliation{Centre for Quantum Technologies, National University of Singapore, 3 Science Drive 2, 117542, Singapore}%

\date{\today}

\begin{abstract}
In a recent work [\href{https://journals.aps.org/prl/abstract/10.1103/PhysRevLett.120.170502} {Phys. Rev. Lett. 120,
170502 (2018)}], Pallister \textit{et al.} proposed an optimal strategy to verify non-maximally entangled two-qubit
pure states under the constraint that the accessible measurements being locally projective and non-adaptive. Their nice
result leads naturally to the question: What is the optimal strategy among general LOCC measurements? In this Letter,
we answer this problem completely for two-qubit pure states. To be specific, we give the optimal strategy for each of
the following available classes of measurements: (i) local operations and one-way classical communication (one-way
LOCC) measurements; (ii) local operations and two-way classical communication (two-way LOCC) measurements; and (iii)
separable measurements. Surprisingly, our results reveal that for the two-qubit pure state verification problem,
two-way LOCC measurements remarkably outperforms one-way LOCC measurements and has the same power as the separable
measurements.
\end{abstract}

\maketitle

\sectionprl{Introduction}On the way to quantum era, quantum devices for generating particular states have been
extensively studied and widely used~\cite{monz201114,wang2016experimental,song201710,friis2018observation}. As so, it
becomes necessary to verify that these devices truly work as they are specified reliably and efficiently with
measurements that are accessible. A standard approach is to estimate the output states with quantum state
tomography~\cite{helstrom1967quantum,holevo1982probabilistic,hayashi1997linear,gill2000state,gross2010quantum,sugiyama2013precision,o2016efficient,haah2017sample}.
However, this method is both time-consuming and computationally difficult, even verifying a few-qubit photonic state is
already experimentally challenging~\cite{haffner2005scalable,carolan2014experimental}. Various non-tomographic
approaches have been designed for this
task~\cite{toth2005detecting,flammia2011direct,da2011practical,hayashi2015verifiable,takeuchi2018verification,morimae2017verification,takeuchi2018resource,zhu2018efficient},
using only local measurements. Though these methods achieve considerable efficiency, no optimal method except for the
maximally entangled state~\cite{hayashi2006study,hayashi2006hypothesis,hayashi2008statistical,hayashi2009group} is
known so far.

Given the intrinsic difficulty in state verification, in this Letter we focus on verifying the non-maximally entangled
two-qubit pure states, in hopes of gaining deeper understanding on the verification problem. These states find wide
applications in quantum information theory~\cite[Section 3.3]{gisin2009bell}, making their verification important both
from the theoretical and the experimental points of view. Specially, we construct optimal strategies when different
classes of measurements are available: one-way LOCC measurements, two-way LOCC measurements, and separable
measurements. We find that for the problem under consideration, two-way LOCC measurements achieves the same performance
as that of separable measurements, while outperforms one-way LOCC measurements dramatically.

Before presenting the results, we review the notations. We denote by $\ket{+}\equiv(\ket{0}+\ket{1})/\sqrt{2}$ and
$\ket{-}\equiv(\ket{0}-\ket{1})/\sqrt{2}$ the eigenstates of the Pauli $X$ operator, by
$\ket{\top}\equiv(\ket{0}+i\ket{1})/\sqrt{2}$ and $\ket{\bot}\equiv(\ket{0}-i\ket{1})/\sqrt{2}$ the eigenstates of the
Pauli $Y$ operator. When measuring a qubit with a Pauli operator, the outcome is written as $(-1)^{i}$ where
$i\in\{0,1\}$. We denote by $\cH$ the two-qubit composite system and by $\1$ the identity operator on $\cH$. We say a
positive operator $T$ with $0\leq T\leq \1$ a one-way LOCC (local operations and only one-way classical communication)
POVM element on $\cH$ if the two-outcome POVM $\{T,\1-T\}$ can be implemented by one-way LOCC. We also define a two-way
LOCC (local operations and two-way classical communication) POVM element and a separable POVM element in the same way
by using the two-way LOCC and the separable operations, respectively. Interested readers might refer
to~\cite{hayashi2016quantum} for details on these operations. We write the set of one-way LOCC from Alice to Bob,
one-way LOCC from Bob to Alice, two-way LOCC, and separable POVM elements as $\cT_{\rightarrow}$, $\cT_{\leftarrow}$,
$\cT_{\leftrightarrow}$, and $\cT_ {\opn{sep}}$. These classes satisfy the relation
$\cT_{\rightarrow}(\cT_{\leftarrow})\subseteq\cT_{\leftrightarrow}\subseteq\cT_{\opn{sep}}$. The condition $T\in\cT_c$
is equivalent to the condition $\1-T\in\cT_c$, where $c\in\{\rightarrow,\leftarrow,\leftrightarrow,\opn{sep}\}$. For a
positive operator $\Omega$ on $\cH$, $\lambda_i(\Omega)$ denotes the $i$-th eigenvalue of $\Omega$ and
$\lambda_i^\downarrow(\Omega)$ denotes the $i$-th largest eigenvalue of $\Omega$, where $i=1,2,3,4$.

\sectionprl{Two-qubit pure state verification}Consider a quantum device that is designed to produce the two-qubit pure
state
\begin{align}
    \ket{\Psi} = \sqrt{1 - \lambda}\ket{00} + \sqrt{\lambda}\ket{11},
\end{align}
where $\lambda\in[0,1/2]$. However, it might work incorrectly and actually outputs states $\sigma_1,
\sigma_2,\cdots,\sigma_N$ in $N$ runs. It is promised the fidelity $\bra{\Psi}\sigma_j\ket{\Psi}$ is either $1$ or
satisfies $\bra{\Psi}\sigma_j\ket{\Psi}\leq 1 - \epsilon$ for all $j$ for some $\epsilon > 0$. The task is to determine
which is the case. The conclusion is useful if we assume the next state $\sigma_{N+1}$ has the same behavior as the
previous ones.

To achieve this task, we perform two-outcome measurements from a set of accessible measurements to test the state. Each
two-outcome measurement $\{T_l, 1-T_l\}$ is specified by an operator $T_l$, which corresponds to passing the test, and
is performed with probability $p_l$. We require that the target state $\ket{\Psi}$ always passes the test, that is,
$T_l\ket{\Psi}=\ket{\Psi}$ for all $T_l$. In the bad case, the maximal probability that $\sigma_j$ passes the test is
given by~\cite{pallister2018optimal,zhu2018efficient}
\begin{align*}
  \max_{\bra{\Psi}\sigma_j\ket{\Psi}\leq 1-\epsilon }\tr\left(\Omega \sigma_j\right)
= 1 - \left[1-\lambda_2^\downarrow(\Omega)\right]\epsilon,
\end{align*}
where $\Omega=\sum_l p_l T_l$ is called an strategy. After $N$ runs, $\sigma_j$ in the bad case can pass all tests with
probability at most $[1-[1-\lambda_2^\downarrow(\Omega)]\epsilon]^N$. Hence to achieve confidence $1 - \delta$, it
suffices to take~\cite{pallister2018optimal}
\begin{equation}\label{eq:NumTest}
N \geq \frac{\ln\delta}{\ln[1-[1-\lambda_2^\downarrow(\Omega)]\epsilon]}
\approx \frac{1}{[1-\lambda_2^\downarrow(\Omega)]\epsilon}\ln\frac{1}{\delta}.
\end{equation}

The optimal strategy is obtained by minimizing the second largest eigenvalue $\lambda_2^\downarrow(\Omega)$. If there
is no restriction on the accessible measurements, the optimal strategy is given by the measurement
$\{\proj{\Psi},\1-\proj{\Psi}\}$, under which $\Omega=\proj{\Psi}$, $\lambda_2^\downarrow(\Omega)=0$, and
$N\approx\epsilon^{-1}\ln\delta^{-1}$. This efficiency cannot be improved if collective measurements are
allowed~\cite{zhu2018efficient}. However, it is difficult to perform such measurements experimentally when $\ket{\Psi}$
is entangled. It is thus meaningful to devise efficient (or even optimal) strategies based on measurements satisfying
reasonable constraints. Owari and Hayashi~\cite{owari2008two} studied the case where the incorrect states are the
maximally mixed state, with the target to minimize the trace of $\Omega$. They derived optimal strategies when one-way
LOCC and separable measurements are available, and showed that two-way LOCC measurements remarkably improves the
performance compared to one-way LOCC measurements. Recently, Pallister, Linden, and
Montanaro~\cite{pallister2018optimal} proposed an optimal strategy $\Omega_{\opn{PLM}}$ to verify $\ket{\Psi}$, under
the constraint that the accessible measurements must be locally projective and non-adaptive. The strategy
$\Omega_{\opn{PLM}}$~\cite{pallister2018optimal_supp} has the second largest eigenvalue
\begin{equation}\label{eq:Omega-PLM}
  \lambda_2^\downarrow\left(\Omega_{\opn{PLM}}\right) 
  = \frac{2 + 2\sqrt{\lambda(1-\lambda)}}{4 + 2\sqrt{\lambda(1-\lambda)}}.
\end{equation}
Note that the set of accessible measurements in~\cite{pallister2018optimal} forms a strict subset of
$\cT_{\rightarrow}$.

These interesting results lead to the question: What is the optimal strategy when general LOCC measurements,
adaptive choices of local measurements are available? In this paper, we investigate this problem comprehensively. We
derive optimal strategies for verifying $\ket{\Psi}$ when the following different classes of measurements are
available: $\cT_{\rightarrow}$, $\cT_{\leftarrow}$, $\cT_{\leftrightarrow}$, and $\cT_{\opn{sep}}$. In the following,
we say a strategy $\Omega$ is in $\cT_c$ and written $\Omega\in\cT_c$ if $\Omega=\sum_lp_lT_l$ and $T_l\in\cT_c$ for
all $l$, a strategy $\Omega$ is \textit{optimal} in $\cT_c$ if $\Omega\in\cT_c$ and for arbitrary
$\Omega^\prime\in\cT_c$ satisfying $\Omega'\ket{\Psi}=\ket{\Psi}$,
$\lambda_2^\downarrow(\Omega)\leq\lambda_2^\downarrow(\Omega^\prime)$.

Here we discuss some general properties of arbitrary strategy $\Omega$.
Consider the product of local unitaries $U_\theta\otimes U_{-\theta}$, where $U_\theta=\proj{0}+e^{i\theta}\proj{1}$
and $\theta\in[0,2\pi]$. Let $\ket{\Psi^\bot} = \sqrt{\lambda}\ket{00}-\sqrt{1-\lambda}\ket{11}$, then
$\{\ket{\Psi},\ket{\Psi^\bot},\ket{01},\ket{10}\}$ are the four eigenstates of $U_\theta\otimes U_{-\theta}$. Using
this property we can simplify the form of $\Omega$ by \textit{averaging}, where the averaged strategy is defined as
\begin{align*}
\Omega_a 
:=& \frac{1}{2\pi}\int_{0}^{2\pi} \left(U_\theta\otimes U_{-\theta}\right) \Omega
                                  \left(U_\theta\otimes U_{-\theta}\right)^\dagger d\theta.
\end{align*}
Since the second largest eigenvalues of $\Omega_a$ and $\Omega$ are the matrix norms of $ P^{\perp}\Omega_a P^{\perp}$
and $P^{\perp}\Omega P^{\perp}$ with $P^{\perp}:= \1 - \proj{\Psi}$, we have
$\lambda_2^\downarrow(\Omega_a)\leq\lambda_2^\downarrow (\Omega)$. That is, averaging over $\theta$ cannot make the
strategy worse. As $\1\geq\Omega_a\geq\proj{\Psi}$ and the vectors $\ket{01}$ and $\ket{10}$ of $U_\theta\otimes
U_{-\theta}$ have different eigenvalues from that of $\ket{\Psi}$ and $\ket{\Psi^\bot}$, after averaging $\Omega_a$ can
be expressed as
\begin{equation}\label{eq:averaging}
\Omega_a = \proj{\Psi} + \lambda_2\proj{\Psi^\bot} + \lambda_3\proj{01} + \lambda_4\proj{10}
\end{equation}
for some $\lambda_2,\lambda_3,\lambda_4\in[0,1)$. We also consider the fact that $\ket{\Psi}$ is invariant under qubits
swapping. Using the swapping operation $s$ for the roles of Alice and Bob, the resulting strategy
$\overline{\Omega}_a\coloneqq\frac{1}{2}\Omega_a+ \frac{1}{2} s(\Omega_a)$ has performance at least as good as that of
$\Omega_a$ and admits the form
\begin{equation}\label{eq:swapping}
\overline{\Omega}_a = \proj{\Psi} + \lambda_2\proj{\Psi^\bot} + \lambda_3\left(\proj{01} + \proj{10}\right)
\end{equation}
for some $\lambda_2,\lambda_3\in[0,1)$. We should be careful when using the swapping invariance property, as to
implement the strategy $\overline{\Omega}_a$, an extra step of messaging is required.

The above discussed framework is non-adversarial in the sense that the malicious device produces incorrect states
randomly and independently. One may also consider the adversarial scenario where the malicious device may produce an
arbitrary state $\rho$ on the whole system $\cH^{N+1}$~\cite{zhu2018efficient}. The task is then to ensure that the
reduced state on one system has fidelity larger than $1-\epsilon$ by performing $N$ tests on other systems. We remark
that minimizing the second largest eigenvalue leads the optimization of the strategy even in this scenario.

\sectionprl{Strategy using one-way LOCC measurements}First we propose a strategy in $\cT_{\rightarrow}$. Then we show
it is optimal when only $\cT_{\rightarrow}$ are available. 

Let $\ket{v_{\pm}}=\sqrt{1-\lambda}\ket{0}\pm\sqrt{\lambda}\ket{1}$. Alice performs the X measurement on the target
state and sends outcome $i$ to Bob. If $i=0$, Bob performs measurement $\{\proj{v_+}, \1-\proj{v_+}\}$ and accepts if
the outcome is $v_+$. If $i=1$, Bob performs measurement $\{\proj{v_{-}}, \1-\proj{v_{-}}\}$ and accepts if the outcome
is $v_{-}$. The corresponding POVM element $T_1$ (passing the test) has the form
\begin{align*}
T_1 &= \proj{+}\otimes\proj{v_+} + \proj{-}\otimes\proj{v_{-}}.
\end{align*}
We define other two POVM elements $T_2$ and $T_3$ similarly to $T_1$ but with the X measurement replaced by the Y and Z
measurements on Alice's side, respectively. These two elements read
\begin{align}
T_2 &= \proj{\top}\otimes\proj{w_{-}} + \proj{\bot}\otimes\proj{w_{+}},\nonumber \\
T_3 &= \proj{0}\otimes\proj{0} + \proj{1}\otimes\proj{1},\label{eq:T_3}
\end{align}
where $\ket{w_{\pm}}=\sqrt{1-\lambda}\ket{0}\pm i\sqrt{\lambda}\ket{1}$.
It holds that $T_j\ket{\Psi}=\ket{\Psi}$ and $T_j\in\cT_{\rightarrow}$ for $j=1,2,3$.

The one-way strategy goes as follows. In each round, Alice chooses a measurement from $\{T_1, T_2, T_3\}$ with
\textit{a prior} probability $\{\frac{1-p}{2}, \frac{1-p}{2}, p\}$ to test the state, where $p\in[0,1]$ is a free
parameter. The strategy has the form
\begin{align*}
\Omega_\rightarrow =&\; \frac{1-p}{2}T_1 + \frac{1-p}{2}T_2 + pT_3 \\
=&\; \proj{\Psi} + p \proj{\Psi^\bot} \\
&+ (1-p)\lambda\proj{01} + (1-p)(1-\lambda)\proj{10}.
\end{align*}
Minimizing $\lambda_2^\downarrow(\Omega_\rightarrow)$ w.r.t. $p\in[0,1]$, we get $p=\frac{1-\lambda}{2-\lambda}$ and
\begin{equation*}
\Omega_\rightarrow
= \proj{\Psi} + \lambda_2^{\to}\proj{\Psi^\bot} + \lambda_3^{\to}\proj{01} + \lambda_2^{\to}\proj{10},
\end{equation*}
where $\lambda_2^{\to}=\frac{1-\lambda}{2-\lambda}$ and $\lambda_3^{\to} = \frac{\lambda}{2-\lambda}$. Obviously,
$\Omega_\rightarrow\in\cT_\rightarrow$.

Now we show the optimality of $\Omega_\rightarrow$. Let $|t,s\rangle\coloneqq\sqrt{t}|0\rangle
+e^{is}\sqrt{1-t}|1\rangle$, where $t\in[0,1]$ and $s\in[0,2\pi]$. When a one-way LOCC strategy $\Omega$ detects
$\ket{\Psi}$ with certainty, the strategy is composed of Alice's POVM $ \int 2
\proj{t,s} P_{TS}(dt ds)$ with some probability distribution $P_{TS}$ and Bob's two-outcome measurements
$\{\proj{t,s,B} ,\1- \proj{t,s,B}\}$, where $ |t,s,B\rangle $ is the normalized vector of $\sqrt{t(1-\lambda)}|0\rangle
+e^{-is }\sqrt{(1-t)\lambda}|1\rangle $. Then, the strategy $\Omega$ is written as
\begin{align}
\Omega  = 2 \int \proj{t,s} \otimes \proj{t,s,B} P_{TS}(d t d s).\label{vari}
\end{align}
Following the averaging argument in Eq.~\eqref{eq:averaging}, we get $\Omega_a$, obtained from $\Omega$. For the
analysis of $\Omega_a$, we treat the variable $t$ in Eq.~\eqref{vari} as the random variable $T$ subject to $P_{TS}$,
and focus on the expectation $\bE_T$ under the marginal distribution $P_{T}$. To guarantee that Alice's measurement in
$\Omega_a$ is a POVM, $\bE_T[T]=\frac{1}{2}$ needs to hold. In the Supplemental Material~\cite{supplemental} we show
that $\Omega_a$ satisfies Eq.~\eqref{eq:averaging} with
\begin{align*}
  \lambda_2(\Omega_a) =  1 - \Xi,\;
  \lambda_3(\Omega_a) = \Xi\lambda,\; \lambda_4(\Omega_a) = \Xi(1-\lambda),
\end{align*}
where $\Xi\coloneqq 2\bE_T\left[\frac{T(1-T)}{T+\lambda-2\lambda T}\right]\geq0$. As
$\lambda_3(\Omega_a)\leq\lambda_4 (\Omega_a)$, $\lambda_2^\downarrow(\Omega_a)$ is minimized when
$\lambda_2(\Omega_a)=\lambda_4(\Omega_a)$. Solving the equation, we get $\Xi = \frac{1}{2-\lambda}$ and
$\lambda_2^\downarrow(\Omega_a)=\frac{1-\lambda}{2-\lambda}$. This concludes the optimality of $\Omega_\rightarrow$.

Switching the role between Alice and Bob, we get a symmetric version $\Omega_\leftarrow$ of $\Omega_\rightarrow$.
Consider the new strategy $\widehat{\Omega}_{\leftrightarrow}=(\Omega_\rightarrow+\Omega_\leftarrow)/2$. Minimizing the
second largest eigenvalue of $\widehat{\Omega}_{\leftrightarrow}$ w.r.t. $p$ gives
\begin{align}\label{eq:Omega-two-way-two-step}
\widehat{\Omega}_{\leftrightarrow} 
&= \proj{\Psi} + \frac{1}{3}\left(\1-\proj{\Psi}\right).
\end{align}
This two-way two-step strategy outperforms $\Omega_\rightarrow$ in the small regime of $\lambda$. More details on
$\widehat{\Omega}_{\leftrightarrow}$ can be found in the Supplemental Material~\cite{supplemental}.

\sectionprl{Strategy using two-way LOCC measurements}First we describe two measurements both detecting $\ket{\Psi}$
correctly. They are inspired by the two-way LOCC test given in~\cite{owari2008two}. Then we show an appropriate convex
combination of these measurements achieves optimality even if separable measurements are available. In what follows, we
assume $\delta = 1-\sqrt{\frac{\lambda}{1-\lambda}}$ and $p=\frac{\lambda}{1+\sqrt{\lambda(1-\lambda)}}$.

\newcommand{\xratio}{1.2}%
\newcommand{\yratio}{1.0}%
\newcommand{\cmark}{\ding{51}}%
\newcommand{\xmark}{\ding{55}}%
\begin{figure}
\centering
\begin{tikzpicture}
\node[] () at (6*\xratio,-0*\yratio) {Alice};
\node[] () at (6*\xratio,-1*\yratio) {Bob};
\node[] () at (6*\xratio,-2*\yratio) {Alice};
\node[] (M0) at (0.5*\xratio,0*\yratio) {$\delta\proj{0}$};
\node[] (M1) at (3.5*\xratio,0*\yratio) {$(1-\delta)\proj{0}+\proj{1}$};
\node[] (M0_B_0) at (0*\xratio,-1*\yratio) {$\proj{0}$};
\node[] (M0_B_1) at (1*\xratio,-1*\yratio) {$\proj{1}$};
\node[] (M1_B_0) at (2.5*\xratio,-1*\yratio) {$\proj{+}$};
\node[] (M1_B_1) at (4.5*\xratio,-1*\yratio) {$\proj{-}$};
\node[] (M1_B_0_A_0) at (2*\xratio,-2*\yratio) {$\sigma_0$};
\node[] (M1_B_0_A_1) at (3*\xratio,-2*\yratio) {$\1-\sigma_0$};
\node[] (M1_B_1_A_0) at (4*\xratio,-2*\yratio) {$\sigma_1$};
\node[] (M1_B_1_A_1) at (5*\xratio,-2*\yratio) {$\1-\sigma_1$};
\node[] (M0_B_0_acc) at (0*\xratio,-2*\yratio) {\cmark};
\node[] (M0_B_1_rej) at (1*\xratio,-2*\yratio) {\xmark};
\node[] (M1_B_0_A_0_acc) at (2*\xratio,-3*\yratio) {\cmark};
\node[] (M1_B_0_A_1_rej) at (3*\xratio,-3*\yratio) {\xmark};
\node[] (M1_B_1_A_0_acc) at (4*\xratio,-3*\yratio) {\cmark};
\node[] (M1_B_1_A_1_rej) at (5*\xratio,-3*\yratio) {\xmark};
\draw[->,thin] (M0.south) to [out=270,in=90] (M0_B_0.north);
\draw[->,thin] (M0.south) to [out=270,in=90] (M0_B_1.north);
\draw[->,thin] (M1.south) to [out=270,in=90] (M1_B_0.north);
\draw[->,thin] (M1.south) to [out=270,in=90] (M1_B_1.north);
\draw[->,thin] (M1_B_0.south) to [out=270,in=90] (M1_B_0_A_0.north);
\draw[->,thin] (M1_B_0.south) to [out=270,in=90] (M1_B_0_A_1.north);
\draw[->,thin] (M1_B_1.south) to [out=270,in=90] (M1_B_1_A_0.north);
\draw[->,thin] (M1_B_1.south) to [out=270,in=90] (M1_B_1_A_1.north);
\draw[->,thin] (M0_B_0.south) to [out=270,in=90] (M0_B_0_acc.north);
\draw[->,thin] (M0_B_1.south) to [out=270,in=90] (M0_B_1_rej.north);
\draw[->,thin] (M1_B_0_A_0.south) to [out=270,in=90] (M1_B_0_A_0_acc.north);
\draw[->,thin] (M1_B_0_A_1.south) to [out=270,in=90] (M1_B_0_A_1_rej.north);
\draw[->,thin] (M1_B_1_A_0.south) to [out=270,in=90] (M1_B_1_A_0_acc.north);
\draw[->,thin] (M1_B_1_A_1.south) to [out=270,in=90] (M1_B_1_A_1_rej.north);
\end{tikzpicture}
\caption{The two-way measurement $\{T_1^{A\to B},\1-T_1^{A\to B}\}$. Alice first performs measurement
        $\{\delta\proj{0}, (1-\delta)\proj{0}+\proj{1}\}$ and sends outcome to Bob. Conditioned on the outcome, Bob
        adopts different measurements on his post-measurement state and sends outcome to Alice if necessary. Alice then
        performs the corresponding two-outcome measurement to detect the final state she holds.}
\label{fig:measurement-I}
\end{figure}
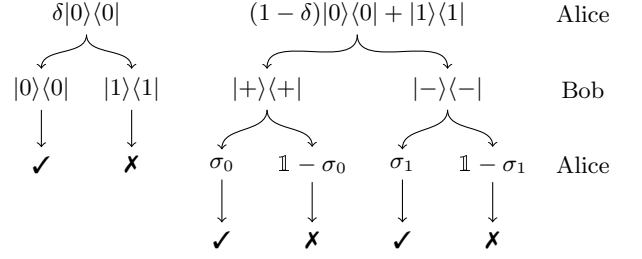

Consider the following measurement procedure:
\begin{enumerate}[leftmargin=*]
\item Alice performs measurement $\{M_0\equiv\delta\proj{0}, M_1\equiv(1-\delta)\proj{0}+\proj{1}\}$ and sends
      the measurement outcome $M_i$ to Bob.
\item Conditioning on $i$, Bob does the following. 
      If $i=0$, Bob performs Z measurement and accepts when the outcome is $0$.
      If $i=1$, Bob performs X measurement and sends outcome $j\in\{0,1\}$ to Alice.
\item Conditioning on $j$, Alice performs measurement $\{\sigma_{j}, \1 - \sigma_{j}\}$ to check the state she
      holds, where $\sigma_{j}$ is the post-measurement state on Alice's system when the input state is
      $\ket{\Psi}$. If she detects $\sigma_{j}$, she accepts.
\end{enumerate}
The corresponding POVM element $T_1^{A\to B}$ (passing the test) has the form
\begin{align*}
T_1^{A\to B} 
=&\; \delta\proj{0}\otimes\proj{0} \\ 
&+ \proj{\widetilde{v}_{+}}\otimes\proj{+} + \proj{\widetilde{v}_{-}}\otimes\proj{-},
\end{align*}
where $\ket{\widetilde{v}_{\pm}}=\sqrt{(1-\delta)A}\ket{0}\pm\sqrt{B}\ket{1}$,
$A=\frac{(1-\lambda)(1-\delta)}{1-\delta+\lambda\delta}$, and $B = \frac{\lambda}{1-\delta+\lambda\delta}$. Note that
$|\widetilde{v}_{\pm}\rangle$ are not normalized. See Fig.~\ref{fig:measurement-I} for illustration of this
measurement. Note that $T_1^{A\to B}\ket{\Psi}=\ket{\Psi}$ and $T_1^{A\to B}\in\cT_{\leftrightarrow}$. The superscript
$A\to B$ of $T_1$ indicates that $T_1$ begins with Alice sending outcome to Bob. A symmetric element $T_1^{B\to A}$ is
obtained by switching the role between Alice and Bob. We define another POVM element $T_2^{A\to B}$ analogous to
$T_1^{A\to B}$ but with the X measurement replaced by the Y measurement on Bob's side, which reads
\begin{align*}
T_2^{A\to B} 
=&\; \delta\proj{0}\otimes\proj{0} \\ 
&+ \proj{\widetilde{w}_{-}}\otimes\proj{\top} + \proj{\widetilde{w}_{+}}\otimes\proj{\bot},
\end{align*}
where $\ket{\widetilde{w}_{\pm}} = \sqrt{(1-\delta) A}\ket{0} \pm i\sqrt{B}\ket{1}$. By construction, $T_2^{A\to
B}\ket{\Psi}=\ket{\Psi}$ and $T_2^{A\to B}\in\cT_{\leftrightarrow}$.

Our two-way strategy is given by the following procedure. In each round, Alice chooses a measurement from
$\{T_1^{A\to B}, T_2^ {A\to B}, T_1^ {B\to A}, T_2^{B\to A}, T_3\}$ with \textit{a prior} distribution
$\{\frac{1-p}{4},
\frac{1-p}{4}, \frac{1-p}{4},
\frac{1-p}{4}, p\}$ to verify the state, where $T_3$ is defined in Eq.~\eqref{eq:T_3}. If $T_i^{A \to B}$ is chosen,
Alice executes the measurement; If $T_i^{B \to A}$ is chosen, Alice sends notification to Bob to ask Bob executing the
measurement. The corresponding strategy is given by
\begin{align}
  \Omega_{\leftrightarrow} 
&= \frac{p-1}{4}\left(T_1^{A\to B} + T_2^{A\to B} + T_1^{B\to A} + T_2^{B\to A}\right) + pT_3\nonumber \\
&= \proj{\Psi} + \lambda^\ast\left(\1 - \proj{\Psi}\right),\label{eq:Omega-two-way}
\end{align}
where 
\begin{align*}
\lambda^\ast = \frac{\sqrt{\lambda(1-\lambda)}}{1 + \sqrt{\lambda(1-\lambda)}}.
\end{align*}
By construction, $\Omega_\leftrightarrow\ket{\Psi}=\ket{\Psi}$ and $\Omega_\leftrightarrow\in\cT_\leftrightarrow$. 
In the Supplemental Material~\cite{supplemental}, we show how the magic values of $\delta$ and $p$ are chosen.

Our strategy $\Omega_{\leftrightarrow}$ can be implemented by two-way LOCC, using up to three step classical
communication. This makes it possible for experimental implementation. When $\lambda=0$, $\ket{\Psi}=\ket{00}$, whose
optimal strategy is provably given by the measurement $\{\proj{00}, \1-\proj{00}\}$. Our strategy
$\Omega_{\leftrightarrow}$ reduces exactly to this optimal measurement when $\lambda=0$, which means our two-way
strategy is globally optimal for $\ket{00}$. However, all other strategies -- $\Omega_{\opn{PLM}}$,
$\Omega_{\rightarrow}$, $\Omega_{\leftarrow}$, and $\widehat{\Omega}_{\leftrightarrow}$ -- do not share this property.

\sectionprl{Optimality of our two-way strategy}We show the optimality of $\Omega_{\leftrightarrow}$ among strategies
using separable measurements. Since a two-way LOCC measurement is a separable measurement, this optimality also shows
the optimality using two-way LOCC measurements. The proof is divided into two parts: first we prove all optimal
strategies in $\cT_{\opn{sep}}$ are homogeneous, then we construct explicitly an optimal homogeneous strategy in
$\cT_{\opn{sep}}$.

A strategy $\Omega$ for $\ket{\Psi}$ is \textit{homogeneous} if it has the form
\begin{align}\label{eq:homogeneous}
    \Omega = \proj{\Psi} + \delta\left(\1-\proj{\Psi}\right),
\end{align}
where $\delta\in[0,1)$. As examples, the strategies $\widehat{\Omega}_{\leftrightarrow}$ and $\Omega_\leftrightarrow$
are homogeneous. Now we prove that the optimal strategies using separable measurements are always homogeneous.
Following the arguments in Eq.~\eqref{eq:swapping}, we know optimal strategies $\Omega$ in $\cT_{\opn{sep}}$ can always
be written as Eq.~\eqref{eq:swapping} for some $\lambda_2,\lambda_3\in[0,1)$. Assume on the contrary
$\lambda_2\neq\lambda_3$, we then construct homogeneous strategies with smaller second largest eigenvalues than that of
$\Omega$, which in turn violates the optimality of $\Omega$. In~\cite[Theorem 1]{owari2008two}, the authors proposed a
separable test of the form
\[
T_4 = \proj{\Psi} + \sqrt{\lambda(1-\lambda)}\left(\proj{01} + \proj{10}\right).
\]
In case $ \lambda_2 > \lambda_3$, we consider a convex combination between $\Omega$ and $T_4$ such that the combination
is homogeneous. The new strategy has a smaller second largest eigenvalue than that of $\Omega$. We can show the
opposite case in the same way using $T_3$ defined in Eq.~\eqref{eq:T_3} instead of $T_4$.

Now we derive an optimal homogeneous strategy in $\cT_{\opn{sep}}$. We are actually interested in the following
optimization problem:
\begin{align*}
\begin{split}
\min          &\;   \delta \\
\text{s.t.}   &\;   0 \leq \delta \leq 1, 
                    \Omega = \proj{\Psi} + \delta\left(\1 - \proj{\Psi}\right),
                    \Omega \in \cT_{\opn{sep}}.
\end{split}
\end{align*}
As the separability condition is equivalent to the PPT (positive partial transpose) condition for two-qubit
operators~\cite{stormer1963positive,woronowicz1976positive}, this problem can be analytically solved. Denote by
$\Omega^{T_B}$ the partial transpose of $\Omega$ on system $B$. The eigenvalues of $\Omega^{T_B}$ are
\begin{align*}
\lambda_1 &= 1 - \lambda + \lambda\delta,\;
\lambda_2 = \lambda + \delta - \lambda\delta,\\ 
\lambda_3 &= \delta + (1-\delta)\sqrt{\lambda(1-\lambda)},\;
\lambda_4 = \delta - (1-\delta)\sqrt{\lambda(1-\lambda)}.
\end{align*}
As $\lambda_1,\lambda_2,\lambda_3\geq0$ for $\lambda\in[0,1/2]$ and $\delta\in[0,1]$, the condition $\Omega^{T_B} \geq
0$ is then equivalent to $\lambda_4\geq0$, resulting
\begin{align*}
\delta \geq \delta^\ast :=\frac{\sqrt{\lambda(1-\lambda)}}{1 + \sqrt{\lambda(1-\lambda)}}.
\end{align*}
The optimal homogeneous strategy then has the form
\begin{align*}
\Omega_{\opn{sep}} = \proj{\Psi} + \delta^\ast\left(\1 - \proj{\Psi}\right).
\end{align*} 
Together with the fact that optimal strategies in $\cT_{\opn{sep}}$ are always homogeneous, we completely solve the
problem of verifying $\ket{\Psi}$ using separable measurements. What's more, as
$\lambda_2^\downarrow(\Omega_{\leftrightarrow})=\lambda_2^\downarrow(\Omega_{\opn{sep}})$, the optimality can be
achieved by two-way LOCC measurements.

\begin{figure}[!htbp]
  \centering
  \includegraphics[width=0.48\textwidth]{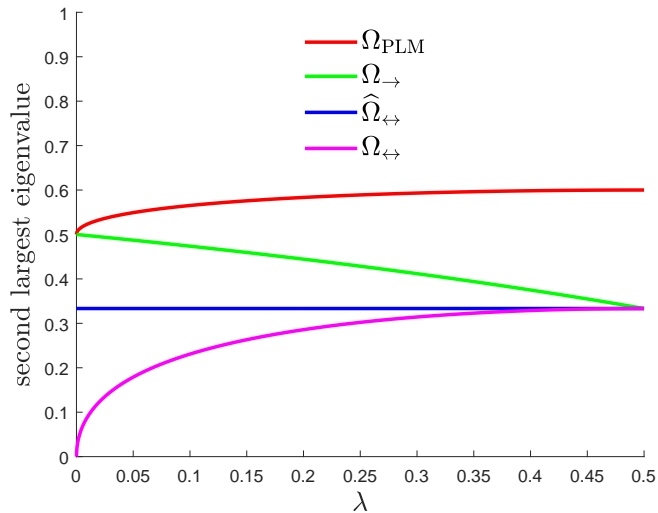}
  \caption{The second largest eigenvalues of our proposed strategies
          -- $\Omega_{\rightarrow}$, $\widehat{\Omega}_{\leftrightarrow}$, and $\Omega_{\leftrightarrow}$ --
          as a function of $\lambda$ (the Schmidt coefficient of $\ket{\Psi}$).
          As comparison, we also plot the second largest eigenvalue of the strategy $\Omega_{\opn{PLM}}$
          proposed in~\cite{pallister2018optimal} (See also Eq.~\eqref{eq:Omega-PLM}).}
  \label{fig:compare-eigenvalues}
\end{figure}

\sectionprl{Comparison}In Fig.~\ref{fig:compare-eigenvalues} we plot the second largest eigenvalues for various
strategies: $\Omega_{\rightarrow}$, $\widehat{\Omega}_{\leftrightarrow}$, $\Omega_{\leftrightarrow}$, and
$\Omega_{\opn{PLM}}$, as a function of $\lambda$, which is the Schmidt coefficient of state $\ket{\Psi}$. One can see
that our proposed strategies give remarkable improvements over $\Omega_ {\opn{PLM}}$, this witnesses the advantage of
adaptivity in state verification: allowing conditional measurements can markedly improve the verification efficiency.
Intuitively, one might expect that the more entangled the $\ket{\Psi}$, the harder to verify it using local
measurements. The two-way strategies $\widehat{\Omega}_{\leftrightarrow}$ and $\Omega_{\leftrightarrow}$ justify this
intuition. However, the one-way strategy $\Omega_{\rightarrow}$, though achieves optimality when $\ket{\Psi}$ is
maximally entangled, has inefficient performance in small regime of $\lambda$, where $\ket{\Psi}$ is less entangled.
This dues to that in the one-way case, the symmetric role between Alice and Bob cannot be utilized. The strict gaps
among $\Omega_{\rightarrow}$, $\widehat{\Omega}_{\leftrightarrow}$ and $\Omega_{\leftrightarrow}$ reveal the power of
classical communication in state verification: with just an extra step of messaging, one can significantly boost the
performance.

\sectionprl{Conclusion}In this Letter, we studied the two-qubit pure state verification problem in depth. We
constructed optimal strategies when the following classes of measurement are accessible: (i) one-way LOCC; (ii) two-way
LOCC; and (iii) separable measurements. Our proposed strategies are dramatically more efficient than all known
candidates based on local measurements and are comparable to the optimal strategy when there is no restriction on the
accessible measurement at all. Our results revealed that for this problem, the two-way LOCC measurements remarkably
outperforms the one-way LOCC measurements and achieves the same performance as the separable measurements. In
principle, the technique used here to construct strategies for verifying two-qubit pure states can be generalized to
pure states of more qubits and higher dimensions. However, it might be rather difficult to get the optimal strategies.

\begin{acknowledgments}
MH is grateful to Dr. Xiao-Dong Yu and Prof. Zhu Huangjun for helpful discussions. 
KW was supported by the National
Natural Science Foundation of China (Grant No. 61300050) and the Program B for Outstanding PhD Candidate of Nanjing
University (Grant No. 201801B047). MH was supported in part by Fund for the Promotion of Joint International Research
(Fostering Joint International Research) Grant No. 15KK0007, a JSPS Grant-in-Aids for Scientific Research (B)
No.16KT0017 and for Scientific Research (A) No.17H01280, and Kayamori Foundation of Information Science Advancement.
\end{acknowledgments}

\bibliographystyle{apsrev4-1}

%

\appendix

\widetext
\clearpage

\setcounter{equation}{0}
\setcounter{figure}{0}
\setcounter{table}{0}
\setcounter{page}{1}
\makeatletter
\renewcommand{\theequation}{S\arabic{equation}}

\begin{center}
\textbf{\large Supplemental Material: Optimal Verification of Two-Qubit Pure States}
\end{center}

The supplemental material is organized as follows. In the first section, we complement the proof that the proposed
strategy using one-way LOCC measurements is optimal. In the second section, we explain the details of the two-way
two-step strategy $\widehat{\Omega}_{\leftrightarrow}$ given in Eq.~(8) of the main text. In the last section, we
show how the magic values of $\delta$ and $p$ are obtained when constructing the optimal strategy
$\Omega_{\leftrightarrow}$ using two-way three-step LOCC measurements.

\section{Optimal strategy using one-way LOCC measurements}\label{app:one-way}

In this section we give more details on the proof that our proposed one-way LOCC strategy is optimal.
As is shown in the main text, a general one-way LOCC strategy for verifying $\ket{\Psi}$ can be written as
\begin{align}\label{app:eq:Omega}
\Omega  = 2 \int \proj{t,s} \otimes \proj{t,s,B} P_{TS}(d t d s).
\end{align}
To analyze $\Omega$, we treat the variable $t$ in Eq.~\eqref{app:eq:Omega} as the random variable $T$ subject to
the marginal distribution $P_{T}$ ans use $\bE_{T}$ to denote the expectation under $P_T$.
The constraint that Alice's measurement in $\Omega$ must be a POVM induces
\begin{equation}\label{eq:POVM}
  \Omega^A = \tr_B\Omega 
= 2\begin{pmatrix} \bE_T\left[T\right] & *
 \\ * & \bE_T\left[1-T\right] \end{pmatrix} = \1.
\end{equation}
Focusing on the diagonal terms, we have the condition $\bE[T] = \frac{1}{2}$. Here, we do not use the condition for the
off-diagonal terms. By letting $\Xi:=2\bE_T\left[\frac{T(1-T)}{D}\right]$ with $D:=T(1-\lambda) + (1-T)\lambda$,
the condition gives the following two relations
\begin{align}
  2\bE_T\left[\frac{T^2}{D}\right](1-\lambda) + \lambda\Xi
= 2\bE_T\left[\frac{T(T(1-\lambda) + (1-T)\lambda) }{D}\right] = 1 
,\label{app:eq:relation1} \\
  2\bE_T\left[\frac{(1-T)^2}{D}\right]\lambda + (1-\lambda)\Xi
= 2\bE_T\left[\frac{T(T(1-\lambda) + (1-T)\lambda) }{D}\right] = 1.\label{app:eq:relation2}
\end{align}
We then get $\Omega_a$ from $\Omega$, using the averaging technique described in the main text. When expressed in the
standard basis, $\Omega_a$ satisfies
\begin{align}\label{app:eq:Omega_a}
\Omega_a 
&= 2
\begin{pmatrix}
  \bE_T\left[\frac{T^2}{D}\right](1-\lambda) 
& 0
& 0
& \bE_T\left[\frac{T(1-T)}{D}\right]\sqrt{\lambda(1-\lambda)} \\
  0
& \bE_T\left[\frac{T(1-T)}{D}\right]\lambda
& 0
& 0 \\
  0 
& 0 
& \bE_T\left[\frac{T(1-T)}{D}\right](1-\lambda)
& 0 \\
  \bE_T\left[\frac{T(1-T)}{D}\right]\sqrt{\lambda(1-\lambda)}
& 0
& 0
& \bE_T\left[\frac{(1-T)^2}{D}\right]\lambda 
\end{pmatrix} \\
&= 
\begin{pmatrix}
1-\lambda \Xi  
& 0
& 0
& \Xi \sqrt{\lambda(1-\lambda)} \\
  0
& \Xi \lambda
& 0
& 0 \\
  0 
& 0 
& \Xi(1-\lambda)
& 0 \\
 \Xi\sqrt{\lambda(1-\lambda)}
& 0
& 0
&
1-(1-\lambda) \Xi  
\end{pmatrix}  \\
&= \proj{\Psi} + \lambda_2\proj{\Psi^\bot} + \lambda_3\proj{01} + \lambda_4\proj{10},
\end{align}
with $\lambda_3 = \bra{01}\Omega_a\ket{01} = \Xi\lambda$, 
$\lambda_4 = \bra{10}\Omega_a\ket{10} = \Xi(1-\lambda)$,
and
\begin{align}
\lambda_2 = \bra{\Psi^\bot}\Omega_a\ket{\Psi^\bot} 
&=  \left[ 1-\lambda\Xi\right]\lambda - 2\lambda(1-\lambda)\Xi
  + \left[ 1-(1-\lambda)\Xi\right](1-\lambda) \\
&= 1 - \left[ \lambda^2 + 2\lambda(1-\lambda) + (1-\lambda)^2\right]\Xi \\
&= 1 - \Xi.
\end{align}

\section{Strategy using two-way two-step LOCC measurements}\label{app:two-way-two-step}

Here we explain in detail the two-way two-step LOCC strategy $\widehat{\Omega}_{\leftrightarrow}$, given in
Eq.~(8) of the main text. We first describe its construction and then prove its optimality when only two-step classical
communication are allowed. Considering the symmetric role between Alice and Bob, we construct from $\Omega_\rightarrow$
a new strategy which outperforms $\Omega_\rightarrow$ in the small regime of $\lambda$. The strategy
$\Omega_\rightarrow$ is implemented by Alice sending measurement outcomes to Bob and Bob performing conditional
measurements. We then get a symmetric version $\Omega_\leftarrow$ of $\Omega_\rightarrow$ by switching the role between
Alice and Bob. The new strategy goes as follows. In each round, Alice first tosses a fair coin, if it is heap up, they
use $\Omega_\rightarrow$; If it is tail up, they use $\Omega_\leftarrow$. The corresponding new strategy then has the
form
\begin{align}
\widehat{\Omega}_{\leftrightarrow} 
&= \frac{1}{2}\Omega_\rightarrow + \frac{1}{2}\Omega_\leftarrow
= \proj{\Psi} + p\proj{\Psi^\bot} + \frac{1-p}{2}\left(\proj{01} + \proj{10}\right).
\end{align}
Minimizing the second largest eigenvalue of $\widehat{\Omega}_{\leftrightarrow}$ w.r.t. $p\in[0,1]$,
we get $p=\frac{1}{3}$ and
\begin{align}
\widehat{\Omega}_{\leftrightarrow} 
&= \proj{\Psi} + \frac{1}{3}\left(\1-\proj{\Psi}\right).\label{Omega-two-way-two-step}
\end{align}

We remark that different from $\Omega_\rightarrow$ and $\Omega_\leftarrow$, $\widehat{\Omega}_{\leftrightarrow}$ must
be implemented by two-way two-step LOCC. This is due to the symmetrization technique we used to construct
$\widehat{\Omega}_{\leftrightarrow}$ from $\Omega_\rightarrow$ and $\Omega_\leftarrow$. Alice and Bob need an extra
step of classical communication to agree on which strategy ($\Omega_\rightarrow$ or $\Omega_\leftarrow$) is used in
current round. Comparing the performance of $\Omega_\rightarrow$ ($\Omega_\leftarrow$) and
$\widehat{\Omega}_{\leftrightarrow}$, one sees the power of classical communication in verification: with just one
extra bit of messaging, $\widehat{\Omega}_{\leftrightarrow}$ outperforms $\Omega_\rightarrow$ ($\Omega_\leftarrow$)
significantly.

We can actually prove that the strategy $\widehat{\Omega}_{\leftrightarrow}$ is the best we can hope when only two-step
classical communication is allowed. Any two-step strategy can be written as a convex combination of one-way LOCC
strategies from Alice to Bob and one-way LOCC strategies from Bob to Alice. In \cite[Theorem 3]{owari2008two} it was
proved that for any one-way LOCC strategy $\Omega$ satisfying $\1 \ge \Omega \ge \proj{\Psi}$, $\tr\Omega\geq2$ holds.
Hence, the second largest eigenvalue of $\Omega$ is no smaller than $\frac{1}{3}$, concluding the optimality of
$\widehat{\Omega}_{\leftrightarrow}$.

\section{Optimization of strategy using two-way LOCC measurements}\label{app:two-way}

When constructing the strategy $\Omega_{\leftrightarrow}$ in the main body, we prefix two magic variables $\delta =
1-\sqrt{\frac{\lambda}{1-\lambda}}$ and $p=\frac{\lambda}{1+\sqrt{\lambda (1-\lambda)}}$.
Here we show that they are actually chosen so that the second largest eigenvalue of $\Omega_{\leftrightarrow}$
is minimized. From now on we assume $\delta\in[0,1]$ and $p\in[0,1]$ are two free parameters to be optimized.
By construction, $\Omega_{\leftrightarrow}$ is given by
\begin{align}
  \Omega_{\leftrightarrow} 
&= \frac{p-1}{4}\left(T_1^{A\to B} + T_2^{A\to B} + T_1^{B\to A} + T_2^{B\to A}\right) + pT_3.
\end{align}
It can be shown that $\Omega_{\leftrightarrow}$ admits the following spectral decomposition:
\begin{align}
  \Omega_{\leftrightarrow} 
&= \proj{\Psi} + \lambda_2(\delta,p)\proj{\Psi^\bot} + \lambda_3(\delta,p)\left(\proj{01} + \proj{10}\right),
\end{align}
where
\begin{align}
\lambda_2(\delta, p) = \frac{p(1-\delta) + \lambda\delta}{1 - \delta + \lambda\delta},\;
\lambda_3(\delta, p) = \frac{ (1-p)\left[\lambda + (1-\lambda)(1-\delta)^2\right]}{2(1 - \delta + \lambda\delta)}.
\end{align}
Our target is to minimize $\lambda_2^\downarrow(\Omega_{\leftrightarrow})$, the second largest eigenvalue of
$\Omega_{\leftrightarrow}$, over the free parameters $\delta$ and $p$ for fixed $\lambda$. This optimization problem
then is given by
\begin{align}
\lambda_2^\downarrow(\Omega_{\leftrightarrow})
:= \min_{\delta\in[0,1], p\in[0,1]}\max\left\{\lambda_2(\delta, p), \lambda_3(\delta, p)\right\}.
\end{align} 
$\lambda_2^\downarrow(\Omega_{\leftrightarrow})$ is minimized for fixed $\lambda$ when the derivatives with respect to
$\delta$ and $p$ vanish. As $\lambda_2(\delta, p)$ is monotonically increasing with $p$ while $\lambda_3(\delta, p)$ is
monotonically decreasing with $p$ in the range $p\in[0,1]$, and $\lambda_2(0,\delta) < \lambda_3(0,\delta)$ for
$\delta\in[0,1]$, $\lambda_2^\downarrow(\Omega_{\leftrightarrow})$ is minimized when $\lambda_2(\delta, p) =
\lambda_3(\delta, p)$. Solving this equation w.r.t. $p$, we get
\begin{align}
p^\ast = \frac{- \lambda\delta^2 + \delta^2 - 2\delta + 1}
              {2\lambda\delta - \lambda\delta^2 + \delta^2 - 4\delta + 3},\;
\lambda_2(\delta) = \lambda_3(\delta) 
= \frac{2\lambda\delta - \lambda\delta^2 + \delta^2 - 2\delta + 1}
        {2\lambda\delta - \lambda\delta^2 + \delta^2 - 4\delta + 3},
\end{align}
where $p^\ast$ is the solution of the equation, which is also the optimal choice of $p$. We now minimize
$\lambda_2(\delta)$ w.r.t. $\delta$. The partial derivative is given by
\begin{align}
  \frac{\partial \lambda_2}{\partial \delta} 
= - \frac{2(1-\lambda)(\delta^2 - 2\delta + \frac{1 - 2\lambda}{1-\lambda})}
          {(2\lambda\delta - \lambda\delta^2 + \delta^2 - 4\delta + 3)^2}
= - \frac{2(1-\lambda)(\delta - \delta_{-})(\delta - \delta_+)}
          {(2\lambda\delta - \lambda\delta^2 + \delta^2 - 4\delta + 3)^2},
\end{align}
where $\delta_{\pm} = 1 \pm \sqrt{\frac{\lambda}{1-\lambda}}$. Solving the equation $\partial \lambda_2/\partial\delta
= 0$ in the range $\delta\in[0,1]$ gives the optimal choice $\delta^\ast=\delta_{-}=1 -
\sqrt{\frac{\lambda}{1-\lambda}}$. Substituting $\delta^\ast$ into $p^\ast$, we get
$p^\ast=\frac{\lambda}{1+\sqrt{\lambda (1-\lambda)}}$ expressed in terms of $\lambda$ solely. Substituting in the
optimal choices of $\delta^\ast$ and $p^\ast$ gives the following optimal strategy
\begin{align}
  \Omega_{\leftrightarrow} 
&= \proj{\Psi} + \frac{\sqrt{\lambda(1-\lambda)}}{1 + \sqrt{\lambda(1-\lambda)}}\left(\1 - \proj{\Psi}\right).
\end{align}
We are done.


\end{document}